\pdfoutput=1   

\documentclass[reprint,superscriptaddress,amsmath,amssymb,amsthm,aps,prx,nofootinbib]{revtex4-2}

\usepackage{graphicx}
\usepackage{dcolumn}
\usepackage{bm}
\usepackage{hyperref}
\usepackage[all]{hypcap}

\usepackage{xcolor}
\usepackage{soul}
\sethlcolor{yellow}

\begin{document}

\preprint{APS/123-QED}

\title{Vibrational, structural, and chemical fingerprints of ion diffusion in crystalline solids}

\author{Gavin Winter}
 \affiliation{Department of Materials Science and Engineering, Massachusetts Institute of Technology, Cambridge, MA 02139, USA}
\author{Juno Nam}
 \affiliation{Department of Materials Science and Engineering, Massachusetts Institute of Technology, Cambridge, MA 02139, USA}
\author{Rafael G\'omez-Bombarelli}
 \email{rafagb@mit.edu}
 \affiliation{Department of Materials Science and Engineering, Massachusetts Institute of Technology, Cambridge, MA 02139, USA}

\date{\today}

\begin{abstract}
Predicting mobile-ion self-diffusivity $D^*$ from molecular dynamics (MD) simulations is essential for identifying promising solid-state electrolytes, but directly simulating ion diffusion is computationally expensive, particularly with high-accuracy machine learning interatomic potentials (MLIPs). Diffusion is a slow, emergent process that requires long trajectories to converge. Thermodynamic properties, by contrast, converge much faster: the enthalpy $h$, vibrational entropy $s_{vib}$, and 2-body, excess configurational entropy $s^{ex}_{2,config}$ can be extracted from comparatively short MD trajectories, and they encode rich information about the free energy landscape from which transport properties like self-diffusivity ultimately arise. Intuitive correlations are discussed between these thermodynamic properties and ion diffusion, motivating a data-driven approach to exploit this link. A simple neural network was trained to predict diffusivity from features computed over short MD trajectories: a vibrational fingerprint (the vibrational density of states, VDOS) and a structural fingerprint (the radial distribution function, RDF), conditioned on chemistry information encoded in the MLIP embedding. This combination allows the model to predict the converged $\log_{10} D^*$ (cm$^2$/s) \textemdash\ normally obtained from significantly longer MD simulations \textemdash\ with a mean absolute error of 0.398 and a Spearman's rank correlation $\rho$ of 0.844.

\end{abstract}

\maketitle


\section{Introduction}
\label{sec:introduction}


Superionic conduction \cite{Boyce1979SuperionicDynamics}, necessary for supporting high current densities in solid-state batteries, is characterized by a room-temperature ionic conductivity on the order of mS/cm and is exhibited by some inorganic crystalline solids \cite{Kamaya2011AConductor,Aono1990IonicPhosphate,Murugan2007FastLi7La3Zr2O12}. Predicting near-room-temperature diffusivity is key for understanding material limitations at the practical operating temperatures of solid-state batteries ($-50$ to $150^{\circ}$C). Strict convergence in ion diffusion statistics can take hundreds of picoseconds to nanoseconds of simulation time, depending on supercell size \cite{He2018StatisticalSimulations,Usler2023ASimulation}. This amounts to days of wall time with state-of-the-art machine learning interatomic potentials (MLIPs) or weeks of wall time with \textit{ab initio} methods for each candidate evaluated.

Self-diffusivity $D^*$ of mobile ions in crystalline solids can be characterized by an Arrhenius dependence on the enthalpy of migration $\Delta h_m$, and a pre-factor $D_0$ that depends on the attempt frequency $\nu_0$ as well as migration entropy $\Delta s_m$.
\begin{align}
    D^* &= D_0 \exp \left( -\frac{\Delta h_m}{k_B T} \right) \label{eqn:diffusivity} \\
    D_0 &\propto \nu_0 \exp \left(\frac{\Delta s_m}{k_B} \right)
\end{align}
For individual crystalline solids, the enthalpic migration barrier $\Delta h_m$ is typically approximated by the 0 K potential energy barrier, which can be determined from the minimum energy path from the nudged elastic band method \cite{Jonsson1998NudgedTransitions}. Such methods require the migration path to be guessed by bond valence site energy or geometric constraints, as demonstrated most recently in a high-throughput manner by Dembitskiy et al. \cite{Dembitskiy2025BenchmarkingMigration}. However, the migration entropy $\Delta s_m$ is more difficult to compute.

One approach to quantify $\Delta s_m$ is to consider the collection of frequencies $\nu_j$ in the ground state and the collection of frequencies $\nu_j'$ in transition state configurations along a migration path corresponding to ion hopping \cite{Dobson1989EntropyHopping,Vineyard1957FrequencyProcesses}.
\begin{equation}
    \Delta s_m = k_B \ln \left( \frac{1}{\nu_0} \frac{ \prod_{j=1}^{3N} \nu_j }{ \prod_{j=1}^{3N-1} \nu_j' } \right)
\end{equation}
In the transition state, there are $3N$ modes (for $N$ atoms in the system) with a single mode that has an imaginary eigenvalue, corresponding to the minimum energy path along the migration coordinate. This approach solely quantifies the vibrational contribution to $\Delta s_m$. Qualitatively, greater degrees of freedom imparted by configurational entropy and site disorder should also increase the magnitude of $\Delta s_m$. The Meyer-Neldel rule, a form of enthalpy-entropy compensation, as applied to solid ion conductors \cite{Dosdale1983TheRegion,Dyre1986ARule,Almond1987TheConductors}, presents an empirical compensation relation that constrains the design of crystalline frameworks with high ionic conductivity since fast ion conduction requires a low $\Delta h_m$ alongside a high $\nu_0$ and $\Delta s_m$. Gelin et al. demonstrated for atomic diffusion in aluminum and silicon that this enthalpy-entropy compensation largely originates from compensation in the vibrational modes of transition states \cite{Gelin2020Enthalpy-entropyPhonons}. Intuitively, creating a transition state requires some bonds to compress (stiffening modes) and other bonds to stretch (softening modes) to accommodate the migrating ion. A lower $\Delta h_m$ (favoring ion diffusion) corresponds to an open bottleneck with less severe lattice distortion. However, in this scenario, the transition state frequencies remain closer to the ground state, consequently decreasing $\Delta s_m$ (penalizing ion diffusion).

Ideally, one could determine $\Delta h_m$, $\Delta s_m$, and $D_0$ independently, to get free energy barriers straight from molecular dynamics (MD), but this would require an automated method for identifying transition states. Transition-state identification methods and rigorous free energy calculations would prove useful for attributing specific characteristics of the solid ion conductor's lattice to fast ion diffusion, but such methods are too computationally expensive and are not readily automated since they require \textit{a priori} knowledge of the migration path or transition state, detracting from their utility as an early-stage screening method. Furthermore, the ground state is often ill-defined in many superionic conductors with liquid-like Li$^+$-ion sublattices, where there are many possible sites and migration pathways for Li$^+$ in a comparatively flat potential energy landscape.

Prior works have associated fast ion transport with polarizable lattices \cite{Bruesch1977BrownianConductors}, low-energy optical phonon modes \cite{Wakamura1997RolesConduction}, and anharmonic lattice dynamics \cite{Ding2020AnharmonicAgCrSe2}, which are all inherently intertwined. Muy et al. first proposed using the band center of the Li$^+$ ion phonon density of states as a descriptor for ion diffusion \cite{Muy2018TuningDynamics,Muy2019High-ThroughputDescriptors,Muy2021PhononIonDynamics}. Other work has further corroborated the correlation between vibrational modes and ion diffusion \cite{Schlem2019ChangingLi6PS5-xSexI,Aghoghovbia2025Machine-learning-assistedConductors,Kim2025AnharmonicConductors,Krauskopf2018ComparingNa3PS4xSex,Xu2022AnharmonicElectrolytes}. Schlem et al. demonstrated how substitution of larger anions in sulfide solid ion conductors drastically improved ion transport, attributing this improvement to a softer lattice \cite{Schlem2019ChangingLi6PS5-xSexI}. Aghoghovbia et al. have shown correlations with Na$^+$ solid ion conductors similar to those originally presented by Muy et al. for Li$^+$ solid ion conductors, correlating band center of Na$^+$ phonon modes with Na$^+$ diffusivity \cite{Aghoghovbia2025Machine-learning-assistedConductors}.

While these works correlated vibrational modes with ion diffusion and provided insight into how vibrational modes can be attributed to specific characteristics of the framework geometry and chemistry, none have directly predicted ion diffusion solely from vibrational modes. These works have focused on computing the phonon density of states from finite displacements, a strictly harmonic approach that depends on $k$-space sampling. By contrast, calculating the vibrational density of states includes the finite-temperature anharmonic effects sampled by MD. A solid ion conductor's vibrational fingerprint is necessary but not sufficient for predicting ion diffusion. Thus, the primary focus of this work is on direct, quantitative prediction of self-diffusivity, using a universal fingerprint that includes vibrational as well as structural and chemical information, rather than focusing on attribution of characteristics (solid ion conductor geometry and chemistry) that lead to fast ion diffusion for a specific material system. A novel metric for quantifying the local interaction between mobile and framework ions' vibrational modes is also presented, as it relates to ion diffusion.

\section{Results}
\label{sec:results}

\subsection{Correlating global vibrational characteristics with ion diffusion}

The total vibrational density of states (VDOS, $g(\nu)$) can be decomposed by element to obtain the spectra for the mobile ions $j$, integrating the Fourier transform of the velocity autocorrelation function and weighting by mass $m_j$ for all $N$ mobile ions:
\begin{equation}
    g(\nu) = \frac{2 }{3N k_B T} m_{\text{Li}} \int^{\infty}_{-\infty} \left< \sum_{j=1}^{N} \textbf{v}_j(t) \cdot \textbf{v}_j(t + \tau) \right>_t e^{-i 2\pi \nu \tau} d\tau,
    \label{eqn:ft-vacf}
\end{equation}
where $\left< \cdot \right>_t$ denotes an average over the time origin $t$. The Fourier transform of the velocity autocorrelation function presented in Equation \ref{eqn:ft-vacf} reduces to the Green-Kubo formalism for self-diffusivity $D^*$ when $\nu = 0$.

\begin{figure*}[hbtp]
    \centering
    \includegraphics[width=\textwidth]{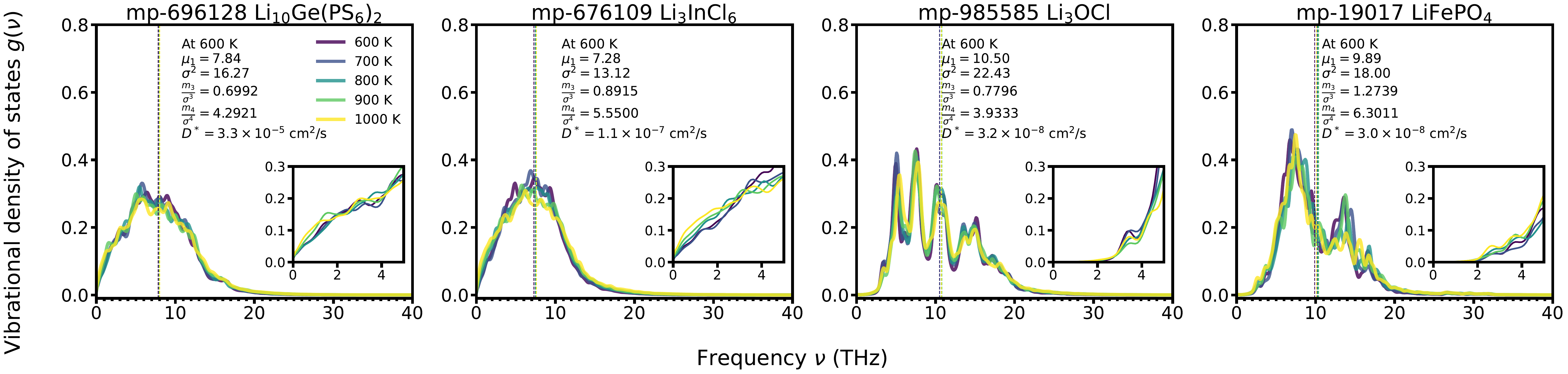}
    \caption{ Vibrational density of states (VDOS) for Li$^+$ in two representative superionic conductors (\textit{left}) and two poor ion conductors (\textit{right}) at a range of temperatures. Dashed line on the vibrational density of states plot depicts the band center $\mu_1$. The annotated self-diffusivity value $D^*$ is from a 100-ps trajectory at 600 K. \label{fig:vdos_temp}}
\end{figure*}

Figure \ref{fig:vdos_temp} shows the VDOS for Li$^+$ in two experimentally-known superionic conductors (mp-696128 Li$_{10}$Ge(PS$_6$)$_2$ and mp-676109 Li$_3$InCl$_6$) and two slower solid ion conductors (mp-985585 Li$_3$OCl and mp-19017 LiFePO$_4$). The VDOS for the superionic conductors is less peaked, in favor of a broad distribution of low-frequency modes with their band center $\mu_1$ (marked by dashed lines) shifted to lower frequencies. In the low-frequency region ($\nu < $ 1.2 THz), the VDOS in crystalline solids generally exhibits the following proportionality $g_{solid}(\nu) \propto \nu^2$, in accordance with the Debye model. The slow ion conductors in Figure \ref{fig:vdos_temp} (Li$_3$OCl and LiFePO$_4$) approximately follow quadratic dispersion ($g(\nu) \propto \nu^2$) at low frequency. In contrast, liquids (or liquid-like sublattices in this work), which have diffusion-dominated vibrational modes that lead to translational motion, generally exhibit linear dispersion in the density of states $g_{liquid}(\nu) \propto \nu$ \cite{Jin2024OnEnergies,Zaccone2021UniversalLiquids,Stamper2022ExperimentalLiquids}. The low-frequency dispersion of the VDOS for the superionic conductors in Figure \ref{fig:vdos_temp} (Li$_{10}$Ge(PS$_6$)$_2$ and Li$_3$InCl$_6$) is roughly linear ($g(\nu) \propto \nu$) \textemdash\ characteristic of liquid-like sublattices. This trend is consistent with neutron-scattering measurements showing the mobile ion sublattice of a superionic solid with liquid-like dynamics exhibits this low-frequency $g_{liquid}(\nu) \propto \nu$ dispersion \cite{Ding2025Liquid-likeElectrolyte}. Thus, the general trends here agree well with empirical correlations that have been well-established for solids and liquids, but have only seen limited application to solid ion conductors.

\begin{figure*}[hbtp]
    \centering
    \includegraphics[width=\textwidth]{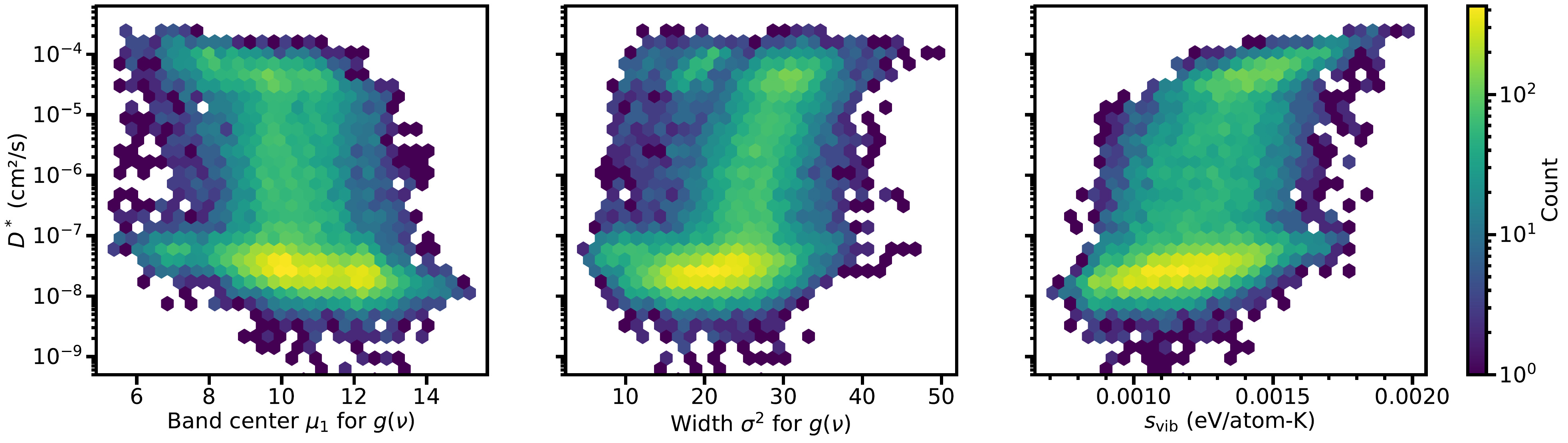}
\caption{ Correlation of self-diffusivity with the moments of the VDOS distribution (left two) as well as the vibrational entropy of the Li$^+$ sublattice (right) for the full set of 4,458 solid ion conductors compiled for all temperatures (600, 700, 800, 900, 1000 K). \label{fig:correlate_diffusivity_vdos_moments}}
\end{figure*}

Given that the transition between the quadratic relation for solids and the linear relation for liquids can be either continuous (as in a glassy-like transition) or discontinuous (as in a first-order thermodynamic transition), Jin et al. generalized the low-frequency regime of the vibrational density of states to $g(\nu) = a(T)\nu^{b(T)} + c(T)$, where $c(T)$ relates to the self-diffusivity in the liquid-phase and generally vanishes for solids \cite{Jin2024OnEnergies}. With sufficient resolution in $g(\nu)$ binning, a simple neural network, as employed in this work, should be able to capture this simple relationship in the hidden layers, incorporating a notion of how liquid-like versus solid-like the mobile ion sublattice is from the low-frequency region of $g(\nu)$.

After computing $g(\nu)$ for many solid ion conductors, the moments of the distribution of vibrational modes can then be correlated with ion diffusion. Ensuring proper normalization $\int_0^\infty g(\nu) d\nu = 1$, the moments can be calculated as follows.
\begin{equation}
    \mu_1 = \bar{\nu} = \int_0^\infty \nu g(\nu) d\nu
    \label{eqn:band_center}
\end{equation}
    

The first raw moment $\mu_1$ of $g(\nu)$ is equivalent to the weighted band center $\bar{\nu}$, capturing the characteristic timescale for vibrations in the system for the Li$^+$ sublattice. The second moment $\sigma^2$ captures the width of the distribution of frequencies in $g(\nu)$.
\begin{equation}
    \sigma^2 = m_2 = \int_0^\infty (\nu - \bar{\nu})^2 g(\nu) d\nu
    \label{eqn:2nd_moment}
\end{equation}
The second moment $m_2$ of $g(\nu)$ describes the average stiffness of the local environment around the Li$^+$ ions, or the curvature of the potential energy surface upon perturbations in Li$^+$ ion coordinates. 


Given the general trends observed empirically for the VDOS of superionic conductors (broad, diffuse, centered at lower frequencies), Figure \ref{fig:correlate_diffusivity_vdos_moments} shows the $D^*$ plotted with respect to each of the $g(\nu)$ moments for 4,458 solid ion conductors at 600, 700, 800, 900, and 1000 K. The higher-order moments $m_3$ and $m_4$ relating to the skewness and kurtosis of $g(\nu)$, respectively, show weak standalone predictive value as descriptors in predicting diffusion, but $D^*$ is plotted versus all of the moments in Figure \ref{fig:correlate_diffusivity_vdos_fullmoments}. As shown in the top row of Figure \ref{fig:correlate_diffusivity_vdos_moments}, lower $\bar{\nu}$ corresponds with higher $D^*$, which is in agreement with the general trend found by Muy et al. \cite{Muy2018TuningDynamics}, while a wider distribution of vibrational modes (higher $m_2$) is associated with a higher $D^*$. For both of these trends, the bi-modal distribution of $D^*$ is quite evident, corresponding to the slow ion conductors and fast ion conductors. The correlations are less clear (or non-existent) for the solid ion conductors with an intermediate $D^*$ value.

The per-atom vibrational entropy $s_{vib}$ of the Li$^+$ sublattice can be defined as follows.
\begin{equation}
    s_{vib} = 3 k_B \int_0^{\infty} g(\nu) \left[ \frac{h \nu}{k_B T}  \frac{1}{e^{\frac{h \nu}{k_B T}} -1} - \ln \left(1 - e^{-\frac{h \nu}{k_B T}} \right) \right] d\nu
    \label{eqn:entropy_migration}
\end{equation}
The $g(\nu)$ would capture \textit{both} the collection of frequencies $\nu_i$ and $\nu_i'$ in Equation \ref{eqn:entropy_migration} as well as the attempt frequency $\nu_0$ embedded in $D_0$ in Equation \ref{eqn:diffusivity}. Since $s_{vib}$ does not distinguish between vibrational contributions near the ground states versus those from the transition states, it does not allow for direct determination of $\Delta s_m$. With unbiased MD, most atomic configurations sampled would correspond with configurations near ground states rather than transition states. The bottom of Figure \ref{fig:correlate_diffusivity_vdos_moments} shows the correlation of $D^*$ versus $s_{vib}$ for all solid ion conductors in the dataset. The correlation of $D^*$ with $s_{vib}$ is better than the correlation of $D^*$ with any of the moments of $g(\nu)$ alone. This is because $s_{vib}$ increases with lower $\bar{\nu}$, broader distribution of $g(\nu)$, and higher population of low-frequency $g(\nu)$ due to greater weighting of low-frequency modes in $s_{vib}$. Thus, $s_{vib}$ essentially captures all the information that the moments capture, as well as the low-frequency dispersion relation.

\subsection{Determining which vibrational frequencies locally contribute to ion diffusion}

\begin{figure*}[hbtp]
\centering{\includegraphics[alt={sample image},width=0.99\textwidth]{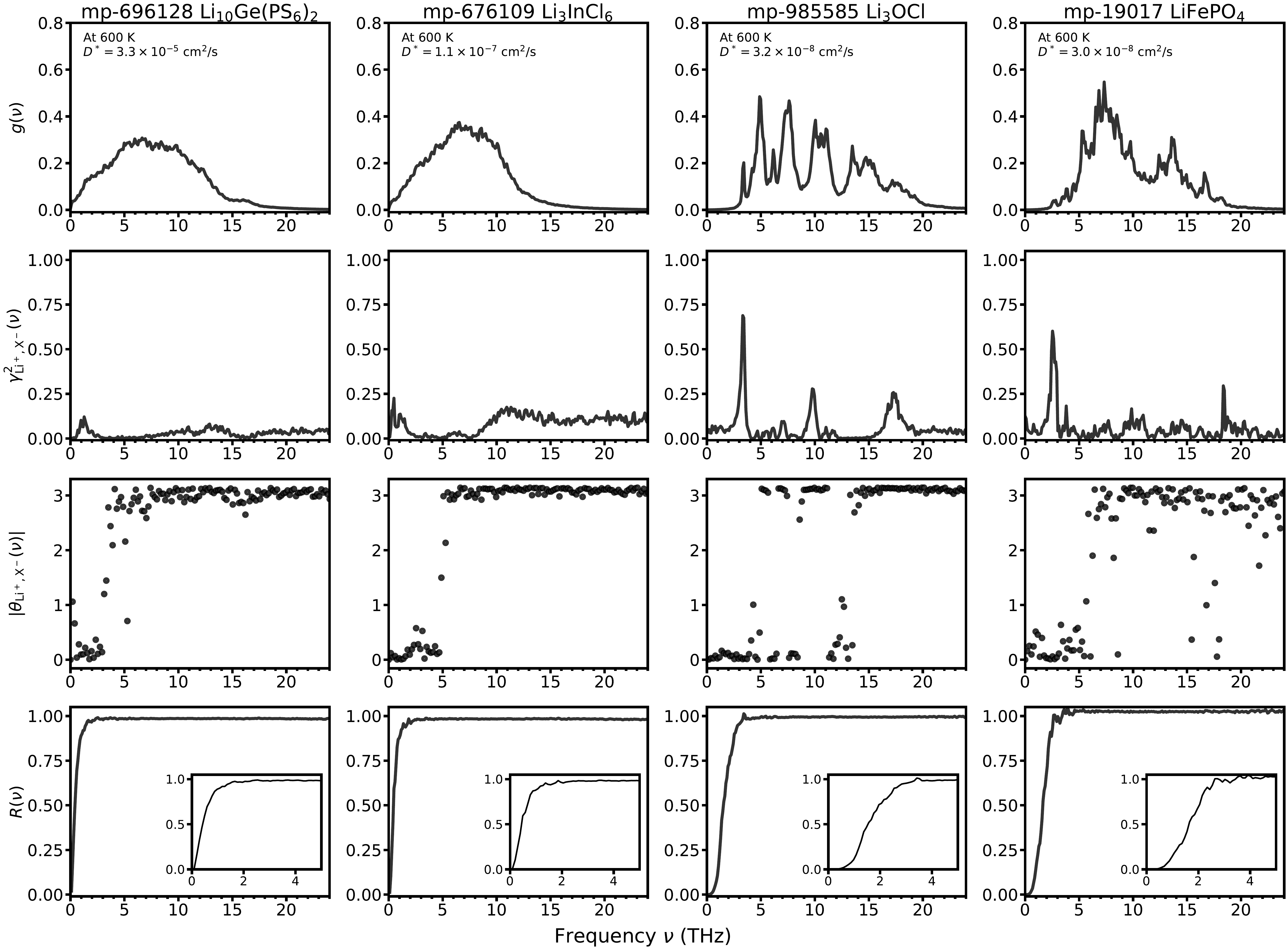}}%
\caption{ Frequency-resolved metrics correlating vibrational motion of mobile Li$^+$ ions and framework ions X$^-$: vibrational density of states $g(\nu)$ (as defined in Equation \ref{eqn:ft-vacf}), coherence $\gamma^2_{\mathrm{Li^+,X^-}}(\nu)$ (as defined in Equation \ref{eqn:coherence}), phase lag $|\theta_{\mathrm{Li^+,X^-}}(\nu)|$, and an inverted stiffness ratio $R(\nu)$ (as defined in Equation \ref{eqn:stiffness}). All spectra presented here are calculated for 600 K. \label{fig:xdoss}}
\end{figure*}

While overlap between the VDOS of Li$^+$ ions and framework ions may indicate there are vibrational modes in the Li$^+$ sublattice and in the framework that coincide at the same frequency, it is not possible to tell solely from the element-decomposed VDOS whether they interact with each other. Even if vibrational modes do interact with each other, they could interact in either a constructive or destructive manner in terms of ion diffusion. A method is presented here for determining whether vibrational modes that occur at a given frequency interact locally and whether framework ions actually impart momentum on Li$^+$ ions. A collective framework ion velocity for the framework ions surrounding a given mobile ion $j$ can be defined as follows:
\begin{equation}
    \mathbf{w}_j(t) = \frac{1}{M_j} \sum_{l \in \mathcal{N}(j)} \mathbf{v}_l(t),
\end{equation}
where $\mathcal{N}(j)$ is the set of $M_j$ framework neighbors of mobile ion $j$, and $\mathbf{v}_l$ is the velocity of a given framework ion $l$.


Analogous to how $g(\nu)$ is represented by a power spectral density that is the Fourier transform of the velocity autocorrelation function for the Li$^+$ sublattice, a cross-spectral density was computed with the Fourier transform of the velocity cross-correlation function between all $N$ mobile ions $j$ and $M_j$ neighboring framework ions $l$ in the Li$^+$ ion's local atomic environment.
\begin{align}
    &P_{\mathrm{Li^+, X^-}} (\nu) \notag \\
    &= \frac{1}{M_j} \frac{1}{N} \sum_{l \in \mathcal{N}(j)} \int^{\infty}_{-\infty} \left< \sum_{j=1}^{N} \textbf{v}_{j}(t) \cdot \textbf{v}_l(t + \tau) \right>_{t} e^{-i 2 \pi \nu \tau} d\tau
\end{align}
The power spectral density for the Li$^+$ ions and the neighboring framework ions X$^-$ were then computed as follows.
\begin{align}
    P_{\mathrm{Li^+, Li^+}} (\nu) \notag \\
    &= \frac{1}{N} \int^{\infty}_{-\infty} \left< \sum_{j=1}^{N} \textbf{v}_{j}(t) \cdot \textbf{v}_j(t + \tau) \right>_{t} e^{-i 2 \pi \nu \tau} d\tau \\
    P_{\mathrm{X^-, X^-}} (\nu) \notag \\
    &= \frac{1}{N} \int^{\infty}_{-\infty} \left< \sum_{j=1}^{N} \textbf{w}_{j}(t) \cdot \textbf{w}_j(t + \tau) \right>_{t} e^{-i 2 \pi \nu \tau} d\tau
\end{align}
The cross-spectral power density ($P_{\mathrm{Li^+,X^-}}$) is complex-valued, and the power spectral densities ($P_{\mathrm{Li^+,Li^+}}$, $P_{\mathrm{X^-,X^-}}$) are real-valued. By linearity of the Fourier transform $\mathcal{F}[ax(t) + by(t)] = a\mathcal{F}[x(t)] + b\mathcal{F}[y(t)]$ and by linearity of the ensemble average, the cross-spectral density was then computed between each Li$^+$ ion's velocity $\mathbf{v}_j$ and the collective framework ions' velocity $\textbf{w}_j$. This is equivalent to taking the average cross-spectral density between \textit{each} Li$^+$ ion's velocity $\mathbf{v}_j$ and \textit{each} neighboring framework ion's velocity $\textbf{v}_l$, but it reduces the computational cost.
\begin{equation}
    \frac{1}{M_j} \frac{1}{N} \sum_{l \in \mathcal{N}(j)} \sum_{j=1}^N  P(\textbf{v}_{j}, \textbf{v}_{l}) = \frac{1}{N} \sum_{j=1}^N P(\textbf{v}_{j}, \textbf{w}_j)
\end{equation}
Only neighbors within a 5~\AA\ cutoff radius around each Li$^+$ ion were considered because neighboring ions far away from the central ion would be less correlated with the central ion's vibrations and dilute the signal-to-noise ratio. Coherence $\gamma^2_{\mathrm{Li^+,X^-}}$ was defined as follows.
\begin{equation}
    \gamma^2_{\mathrm{Li^+,X^-}} (\nu) = \frac{P_{\mathrm{Li^+, X^-}}(\nu) \cdot P_{\mathrm{Li^+, X^-}}^*(\nu)}{P_{\mathrm{Li^+, Li^+}}(\nu) \cdot P_{\mathrm{X^-, X^-}}(\nu)} \label{eqn:coherence}
\end{equation}
This coherence metric quantifies the degree to which Li$^+$ ions are dynamically coupled to the vibrations of their surrounding framework ions at a particular frequency. In general, it would be expected that for $\gamma^2_{\mathrm{Li^+, X^-}}(\nu) \rightarrow 1$ the Li$^+$ ions are locked into framework vibrational modes, while for $\gamma^2_{\mathrm{Li^+, X^-}}(\nu) \rightarrow 0$ the Li$^+$ ion is decoupled from the framework and moving independently \textemdash\ characteristic of the liquid-like sublattice in superionic conductors. While the neighbor list is only computed at the beginning of each time window, the mobile ions in a fast ion conductor may completely leave their original sites and have new neighbors by the end of the time window. This is acceptable because it would appear as $\gamma^2_{\mathrm{Li^+, X^-}}(\nu) \rightarrow 0$, since the frequencies of the Li$^+$ ion and the framework ions will no longer be coupled. Thus, decoherence intuitively captures the physics that leads up to ion hopping without necessarily requiring an ion hopping event, potentially reducing the simulation time needed to sufficiently sample ion hopping itself. The second row of Figure \ref{fig:xdoss} shows $\gamma^2_{\mathrm{Li^+, X^-}}(\nu)$ computed from MD trajectories of varying length for two superionic conductors (Li$_{10}$Ge(PS$_6$)$_2$, Li$_3$InCl$_6$) and two slow ion conductors (Li$_3$OCl, LiFePO$_4$) at 600 K. For the superionic conductors, the coherence with the framework is quite low, featuring small peaks around 1-2 THz. This reinforces how low coherence in the low-frequency region suggests the Li$^+$ ions are decoupled from the framework ions and are mobile. In contrast, the slow ion conductors have much larger magnitude peaks in their coherence spectra, with the prominent peaks at slightly higher frequencies around 2-4 THz. As observed in the coherence spectra across many solid ion conductors, there is generally a main peak around 2-5 THz and comparatively little coherence between the Li$^+$ sublattice and the neighboring framework ions above 5 THz, which is consistent with the phase analysis below.

In order to understand whether coherence between the framework ions and the mobile ions is truly synchronized or phase-lagged, the phase information $\theta(\nu) = \mathrm{Arg}(P_{\mathrm{Li^+, X^-}} (\nu))$ can be leveraged.
Values of $\theta (\nu) \rightarrow 0$ represent in-phase vibrations of framework ions and Li$^+$ ions, while $\theta(\nu) \rightarrow \pm \pi$ represents vibrations that are out-of-phase with each other. This phase lag $\theta(\nu)$ between the Li$^+$ sublattice and the framework ions is shown in the third row of Figure \ref{fig:xdoss}. Across the different solid ion conductors in Figure \ref{fig:xdoss}, only frequencies below 5 THz correspond to in-phase ($\theta(\nu) \rightarrow 0$) vibrational modes for the Li$^+$ sublattice and the neighboring framework ions. These in-phase, low-frequency modes can enable coupling between framework ions' vibrational modes and translational motion of Li$^+$ ions. Empirically, for almost all 4,458 solid ion conductors in the dataset, it was observed that most vibrations above 10 THz are out-of-phase ($\theta(\nu) \rightarrow \pm \pi$), which is likely because it is rarer for vibrational modes to be synchronized at higher frequencies.

The aforementioned methods provide a means for determining which vibrational frequencies locally contribute to ion diffusion. While lower frequency vibrational modes are generally considered softer, a spectral relationship for the forces is required to determine whether the vibrations are truly soft. Using the per-atom forces $\textbf{f}_j$ on the mobile ions, a stiffness ratio $R(\nu)$ can be computed to quantify the forces on Li$^+$ during vibrations. The power spectrum for the forces $\Phi(\nu)$ was then used to calculate $R(\nu)$ \cite{Zaccone2023GeneralMotions}.
\begin{equation}
    \Phi(\nu) = \frac{2}{3 N k_B T} \frac{1}{m_j} \int_{-\infty}^{\infty} \left< \sum_{j=1}^{N} \textbf{f}_j(t) \cdot \textbf{f}_j(t + \tau) \right>_t e^{-i 2\pi \nu \tau} d\tau
\end{equation}
\begin{equation}
    R(\nu) = \frac{(2 \pi \nu)^2 g(\nu)}{\Phi (\nu)} \label{eqn:stiffness}
\end{equation}
In a system without external forces, $\Phi(\nu) = (2 \pi \nu)^2 g(\nu)$ (i.e. $R(\nu) = 1$). However, a thermostat was used in all MD simulations shown in this work (see Section \ref{sec:methods}). With the introduction of the thermostat, the ratio $R(\nu)$ can deviate from 1 at low frequencies. Since $\Phi(\nu)$ is calculated from the conservative interatomic forces at each frame in MD, it is unaffected by the thermostat. Unlike the purely conservative interatomic forces, the velocities \textemdash\ which define $g(\nu)$ \textemdash\ capture the complete dynamical trajectory, including the continuous momentum scaling induced by the deterministic thermal bath. Because the numerator ($g(\nu)$) and denominator ($\Phi(\nu)$) are consequently governed by differing force landscapes, the exact mathematical identity ($R(\nu)=1$) is broken. This separation allows $R(\nu)$ to expose the intrinsic anharmonic friction and caging effects of the lattice. The bottom row of Figure \ref{fig:xdoss} shows the stiffness ratio $R(\nu)$ for two superionic conductors and two slow ion conductors. The two fast ion conductors (Li$_{10}$Ge(PS$_6$)$_2$, Li$_3$InCl$_6$) have $R(\nu)$ spectra that approach 1 at smaller $\nu$, as compared with the two slow ion conductors (Li$_3$OCl, LiFePO$_4$). This suggests that the low-frequency modes in the Li$^+$ sublattice may be more active in fast ion conductors. In contrast, the rigid caging in slow ion conductors exerts strong non-linear restoring forces on the Li$^+$ ions, maintaining a large $\Phi(\nu)$ even as $g(\nu)$ diminishes, resulting in a wide deficit where $R(\nu)$ approaches zero. While the absolute numerical value of $R(\nu)$ is perturbed by the thermostat, the relative differences in spectra serve as a probe of the intrinsic migration dynamics between fast versus slow ion conductors.

\subsection{Correlating structural characteristics of the radial distribution function with ion diffusion}

\begin{figure*}[htbp]
    \centering
    \includegraphics[width=\textwidth]{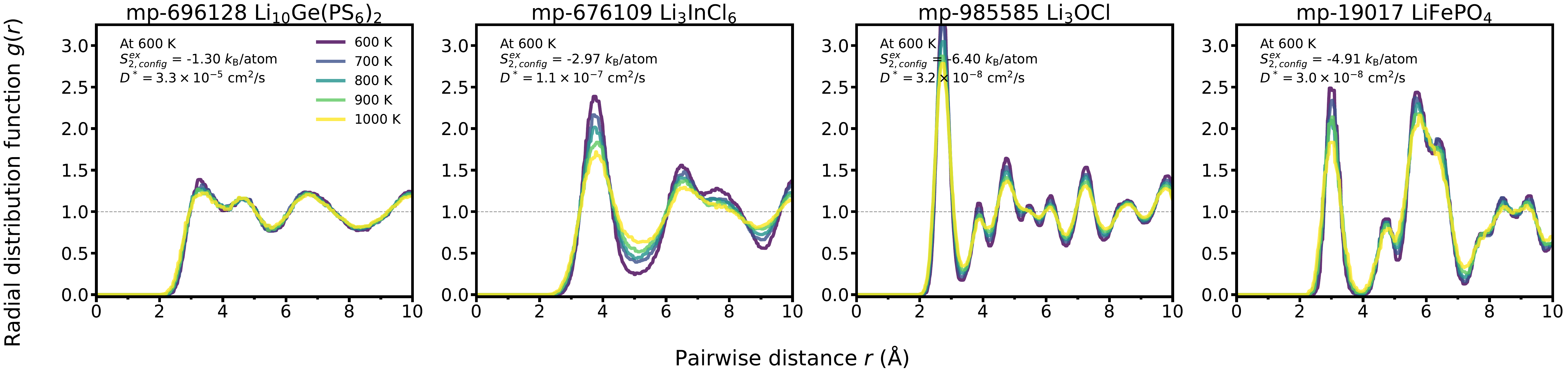}
    \caption{ Radial distribution function (RDF) for Li$^+$-Li$^+$ in two representative superionic conductors (\textit{left}) and two poor ion conductors (\textit{right}) at a range of temperatures. The annotated self-diffusivity value $D^*$ is from a 100-ps trajectory at 600 K. \label{fig:rdf_temp}}
\end{figure*}

\begin{figure}[htbp]
\centering{\includegraphics[alt={sample image},width=0.50\textwidth]{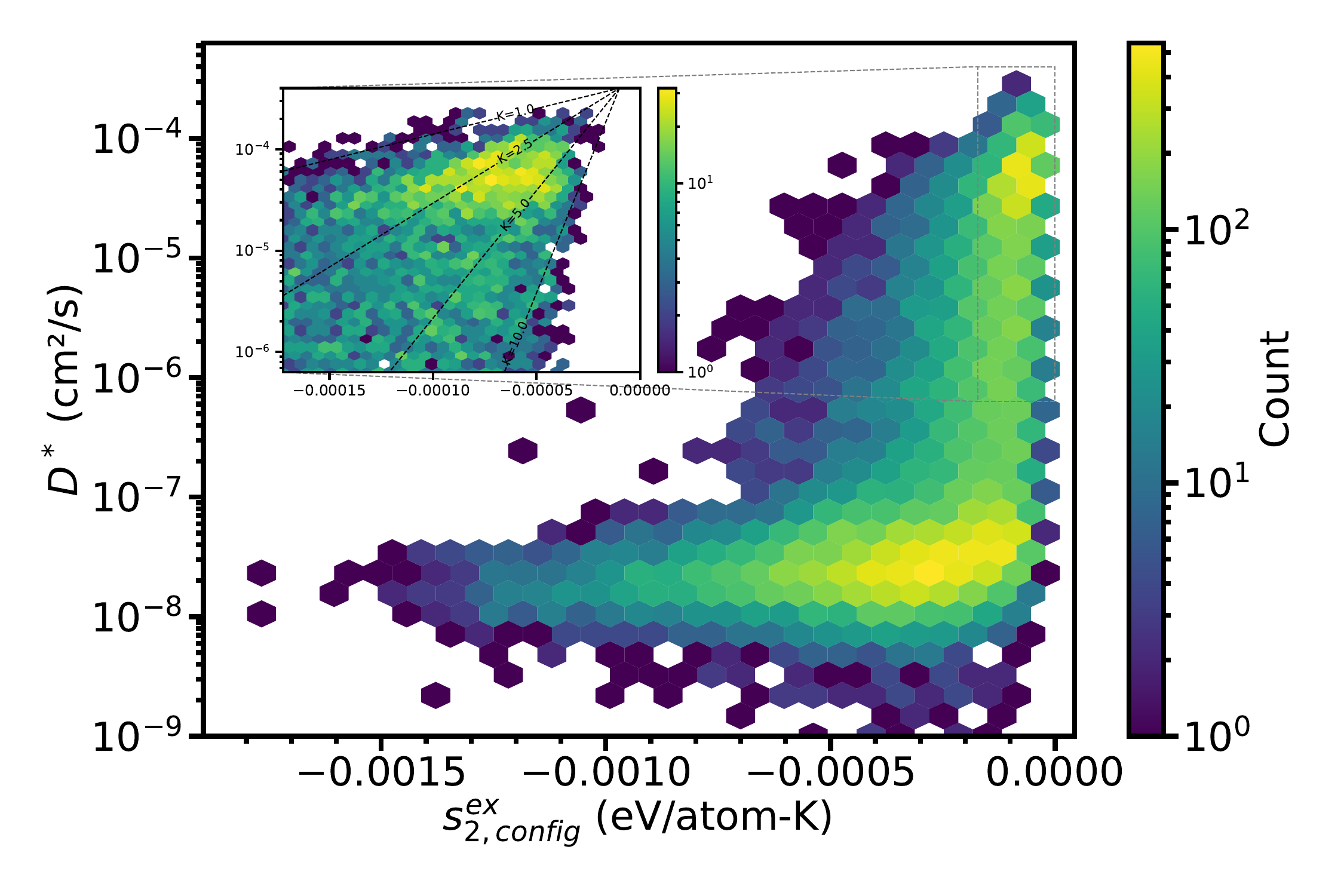}}%
\caption{ Correlation of diffusivity with the 2-body excess configurational entropy for the Li$^+$ sublattice in the full set of 4,458 solid ion conductors for all temperatures (600, 700, 800, 900, 1000 K). The Rosenfeld relation (Equation \ref{eqn:rosenfeld_relation}) is overlaid on the figure inset. \label{fig:correlate_diffusivity_rdf_config}}
\end{figure}

The radial distribution function (RDF, $g(r)$) between two element types (in this case Li$^+$-Li$^+$) can be computed by binning the pairwise distances between all $N$ mobile ions $j$ and normalizing by the number density $\rho$.

\begin{equation}
    g(r) = \frac{1}{4 \pi r^2} \frac{1}{\rho N} \sum_{j}^N \sum_{j' \neq j}^N \delta(r - \lvert\lvert \textbf{x}_j - \textbf{x}_{j'} \rvert\rvert )
    \label{eqn:rdf}
\end{equation}

The pairwise RDF between all the mobile ions determines the excess 2-body configurational entropy of the mobile-ion sublattice \cite{Baranyai1989DirectLiquids,Lazaridis1998InhomogeneousTheory}, relative to the entropy of the ideal gas $s^{id}_{2,config}$ at the same density and temperature. As a result, $s^{ex}_{2,config}$ is by construction negative.

\begin{equation}
    s^{ex}_{2, config} = -2 \pi \rho k_B \int_0^{\infty} [g(r) \ln g(r) - (g(r) - 1)] r^2 dr
    \label{eqn:s_ex2config}
\end{equation}

Assuming the sublattice consisting of the framework anions has comparatively fewer degrees of freedom than the Li$^+$ sublattice in a Li$^+$-ion conducting crystalline solid, this approximates well the 2-body, excess configurational entropy $s^{ex}_{2,config}$. 

With the Rosenfeld relation \cite{Rosenfeld1977RelationSystems,Dzugutov1996AMatter,Dyre2018Perspective:Scaling,Ghaffarizadeh2024AFluids}, $D^*$ in simple liquids such as liquid metals and in the prototypical solid ion conductor $\alpha$-AgI has been shown to exhibit an exponential dependence on $s^{ex}_{2,config}$ \cite{Dzugutov1996AMatter,Hoyt2000TestMetals,Li2005ExcessMetals,Jakse2016ExcessMetals}. Note that $s^{ex}_{2,config}$ presented in this work is a per-atom entropy rather than the dimensionless entropy used in the original Rosenfeld notation \cite{Rosenfeld1977RelationSystems}. 
\begin{equation}
    D^* \propto \exp \left( K \frac{s^{ex}_{2,config}}{k_B} \right) \label{eqn:rosenfeld_relation}
\end{equation}
Liquid metals are strongly interacting systems that may be approximated by Lennard-Jones, hard-sphere fluids, where transport is gated by multi-body caging effects \cite{Dzugutov1996AMatter}. The Rosenfeld relation relies on using the $s^{ex}_{2,config}$ as a measure of the probability of the liquid structure to rearrange itself to let a particle pass. A related interpretation may apply for the mobile sublattice of superionic conductors, in which ion diffusion has been shown to be highly correlated or concerted \cite{He2017OriginConductors}. The Rosenfeld relation presents a necessary but insufficient criterion for ion diffusion, given that the proportionality constant $K$ is related to the collision frequency and given that $s^{ex}_{2,config}$ only explicitly incorporates 2-body interactions \cite{Dzugutov1996AMatter}. While many-body correlation cannot be incorporated without higher-order correlation functions beyond the RDF, the nature and frequency of collisions will be implicitly embedded in the model through the vibrational density of states discussed in the previous section.

Figure \ref{fig:rdf_temp} shows the Li$^+$-Li$^+$ RDF for the same solid ion conductors. The superionic conductor Li$_{10}$Ge(PS$_6$)$_2$ exhibits an RDF characteristic of a liquid-like sublattice, where the peaks corresponding to first, second, and subsequent nearest-neighbor distances are not very large. This indicates the minima in the free energy landscape corresponding to atomic configurations with Li$^+$ occupancy at these sites are not very deep. The probability of finding Li$^+$ ions at distances between the nearest-neighbor sites is nearly as probable as finding them on the sites themselves. In contrast, the slower ion conductors in Figure \ref{fig:rdf_temp} exhibit much more well-defined peaks that correspond with Li$^+$ occupying sites with deep minima, resulting in an immobile, rigid Li$^+$ sublattice. Therefore, it is possible to empirically discern between the liquid-like Li$^+$ sublattice of a superionic conductor and the solid-like Li$^+$ sublattice of a slow ion conductor.

In Figure \ref{fig:correlate_diffusivity_rdf_config}, $D^*$ was plotted against the 2-body, excess configurational entropy $s^{ex}_{2,config}$ calculated by Equation \ref{eqn:s_ex2config} for all 4,458 solid ion conductors. The Rosenfeld relation (Equation \ref{eqn:rosenfeld_relation}) is overlaid on the inset. Fast solid ion conductors are the closest to following the Rosenfeld relation, given that their mobile ion's sublattice is liquid-like, with proportionality constants ($K$) generally ranging from 0.8 to 6.0 (see inset of Figure \ref{fig:correlate_diffusivity_rdf_config}). For liquid metals, the $K$ in the Rosenfeld relation ranges from 0.6 to 0.8 \cite{Li2005ExcessMetals,Rosenfeld1977RelationSystems}. This value found for the solid ion conductors' sublattice is significantly higher, which suggests that ion diffusion is highly sensitive to structural order. This difference may reflect concerted motion and the mobile sublattice's more constrained environment in a solid ion conductor, compared to in a liquid metal. While the proportionality constant for solid ion conductors ranges widely, its lower bound is close to $K=0.8$ \textemdash\ the upper limit for liquid metals. This further reinforces that in approaching the superionic limit the Li$^+$-Li$^+$ interactions in the sublattice may become more akin to the interactions in molten Li metal (experimentally measured $D^* = 3 \times 10^{-4}$ cm$^2$/s at 1000 K \cite{Murday1968SelfDiffusionLithium}). For slower ion conductors ($D^* < 10^{-6}$ cm$^2$/s) that do not feature a liquid-like sublattice, the Rosenfeld relation clearly breaks down. However, for the purpose of predicting ion diffusion across the range of slow and fast ion conductors, the elbow-shaped distribution of $D^*$ and $s^{ex}_{2,config}$ can still act as an envelope function. For instance, below a threshold of $-$0.00075 eV/atom-K, no sample in the dataset exhibits $D^*$ greater than 10$^{-6}$ cm$^2$/s. Thus, $s^{ex}_{2,config}$ was confirmed to serve as a necessary criterion for diffusion in solid ion conductors, but is not by itself sufficient.


\subsection{Vibrational modes and radial distribution function as a universal descriptor for predicting diffusivity}

The goal in using the vibrational modes and radial distribution function as a universal descriptor was to identify properties of the underlying free energy landscape that are very closely related to ion diffusion but converge more quickly in simulation time than ion diffusion statistics, extracting rapidly converging information from a short trajectory. Figure \ref{fig:vdos_rdf_convergence} shows the convergence of the VDOS with respect to trajectory lengths (1, 2, 5, 10, 25 ps). While some peaks and certainly the zero-frequency density of states cannot be resolved well until 25 ps or longer simulation times, the characteristic features for the distribution of frequencies as a whole are already resolved by 5 ps and little new information is gained between 5 ps and 25 ps. From just a 5 ps trajectory, a nearly converged VDOS and RDF can be computed. Convergence in the diffusion statistics of solid ion conductors often requires greater than 100 ps, particularly for slow ion conductors and small supercell sizes \cite{He2018StatisticalSimulations,Usler2023ASimulation}. In contrast, with the exception of the zero-frequency density of states in the VDOS, the simulation time required for convergence of the VDOS and RDF generally does not depend on whether the crystal is a slow or fast ion conductor. Therefore, using the VDOS and RDF provides a means of decoupling simulation time from whether a solid ion conductor has a high diffusivity when predicting ion diffusion.

\begin{figure*}[hbtp]
\centering{\includegraphics[alt={sample image},width=0.75\textwidth]{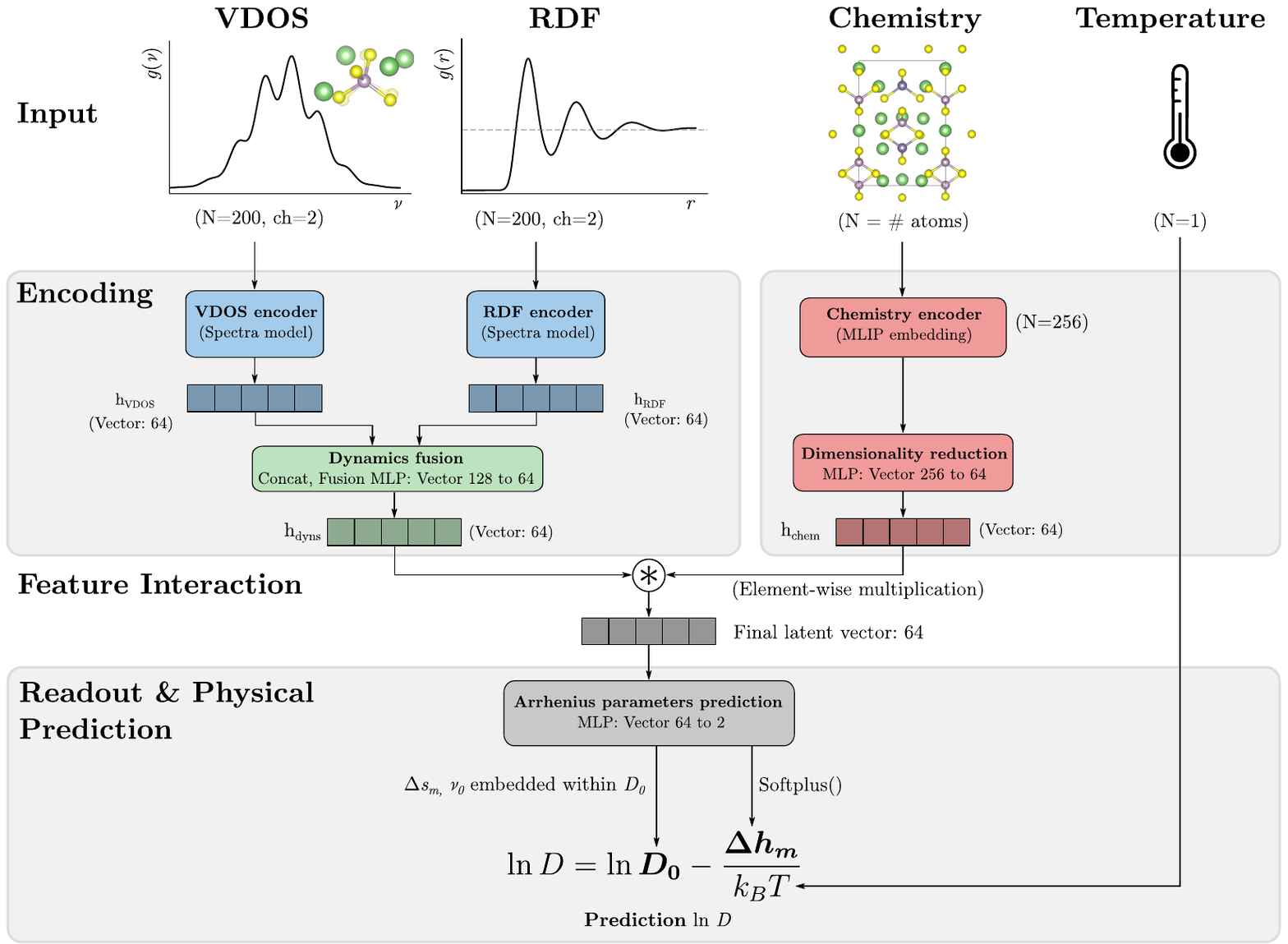}}%
\caption{ Arrhenius-type model used for predicting self-diffusivity. For the spectra model used for encoding the VDOS and RDF, SpectraCNN can be replaced with SimpleNN. \label{fig:arrhenius_architecture_v2}}
\end{figure*}

Neural networks provide a means of utilizing the rich information contained in the VDOS and the RDF, when rigorous physical relationships (constitutive equations) directly connecting both spectra to self-diffusivity do not exist. As shown in Figure \ref{fig:arrhenius_architecture_v2}, an Arrhenius-type model was constructed to predict the self-diffusivity jointly from the VDOS and RDF, while conditioning on the ORB embedding (\texttt{orb-v3-conservative-inf-omat}) \cite{Rhodes2025Orb-v3:Scale}. The ORB embedding of the static (0 K) relaxed crystal contains the chemistry and local environment around each of the Li$^+$ ions that would otherwise only be implicit through the VDOS and RDF. Physically, this Arrhenius-type form was motivated by the fact that: (1) rich vibrational, structural, and chemical information should directly influence different factors in an Arrhenius relationship, and (2) temperature should be allowed to be an explicit parameter since it is known in a simulation. $\Delta h_m$ is informed at least partially by the RDF characteristics and ORB embedding. The pre-factor $D_0$ has $\Delta s_m$ and $\nu_0$ embedded in it, which primarily arises from the VDOS.

Training for each model was performed on a dataset comprising the fingerprint extracted from 5-ps trajectories, using mean square error loss for scalar prediction. The self-diffusivity at 100 ps ($D^*_{\text{100 ps}}$), as evaluated by the mean of the posterior distribution from \texttt{kinisi}, was taken to be the ground truth. To prevent data leakage between the training and test sets from trajectories containing the same crystalline structure at different temperatures, a group shuffle split was employed using material identifiers as groups, ensuring that all trajectories belonging to a given material were strictly segregated into either the training set (17,825) or the test set (4,460). Direct scalar prediction represents scalar prediction taken directly from neural network readout rather than from the Arrhenius-type model shown in Figure \ref{fig:arrhenius_architecture_v2}. For training and inference using the VDOS and RDF spectra, either the SpectraCNN model or a SimpleNN with a single hidden layer was used (detailed in Section \ref{sec:methods}). Adaptations were made to overcome the inherent translational invariance of a convolutional neural network (CNN) such as SpectraCNN, so that interactions across multiple frequency-dependent spectra could be learned. With delta learning by mean-variance estimation \cite{Nix1994EstimatingDistribution}, a correction to the unconverged self-diffusivity $D^*_{\text{5 ps}}$ was learned from the fingerprint extracted from the trajectory at 5 ps (VDOS $g(\nu)$, RDF $g(r)$, and ORB descriptor). The objective for the correction was to match the ground-truth self-diffusivity $D^*_{\text{100 ps}}$. Both the mean and variance of the posterior distribution of $\log_{10} D^*_{\text{100 ps}}$ from \texttt{kinisi} \cite{McCluskey2024AccurateSimulation} were predicted in the objective function. The same group shuffle split was performed in this delta learning task as with the other prediction task.

Table \ref{tab:table1} summarizes the performance of all models tested for predicting $D^*$. Direct scalar prediction performed slightly better than constraining the outputs to the Arrhenius-type model. As a standalone input descriptor, the VDOS performs the best, indicating it provides the richest information regarding diffusion, while the RDF and ORB embedding by themselves are weaker standalone predictors of diffusion. The SpectraCNN model was designed as a more expressive alternative that incorporates multiple channels along the frequency $\nu$ axis (i.e. $g(\nu)$, $R(\nu)$, $\gamma^2_{\mathrm{Li^+,X^-}}(\nu)$). However, the SimpleNN for learning features in the spectra matches or exceeds the performance of the more sophisticated SpectraCNN, indicating that a SimpleNN is sufficient for capturing the relation of 1D spectra to ion diffusion. Delta learning by mean-variance prediction does not improve point-prediction accuracy over scalar prediction. However, mean-variance prediction allows for uncertainty estimation, as will be discussed later.

\begin{table*}[hbtp]
\caption{\label{tab:table1}
Performance in predicting the converged self-diffusivity at 100 ps ($D^*_{\text{100 ps}}$) for each model, using a held-out test set of 4,460 independent 100-ps trajectories at either 600, 700, 800, 900, or 1000 K. VDOS $g(\nu)$, RDF $g(r)$, and $D^*_{\text{5 ps}}$ were obtained from a 5-ps trajectory.}
\begin{ruledtabular}
\renewcommand{\arraystretch}{1.5} %
\begin{tabular}{cccccccc}
Training data & Spectra model & Diffusivity model & MAE ($\log_{10}D^*$ [cm$^2$/s]) & $\rho$ \\
\hline
VDOS & SimpleNN & Direct scalar prediction & 0.485 & 0.835 & \\
VDOS & SpectraCNN & Direct scalar prediction & 0.535 & 0.825 & \\
RDF & SimpleNN & Direct scalar prediction & 0.622 & 0.731 & \\
ORB descriptor & SimpleNN & Direct scalar prediction & 0.621 & 0.737 & \\
VDOS + RDF + ORB descriptor & SimpleNN & Direct scalar prediction &  \textbf{0.392} & \textbf{ 0.858} & \\
VDOS + RDF + ORB descriptor & SimpleNN & \begin{tabular}{@{}c@{}}Scalar prediction \\[-1ex] with Arrhenius-type model \end{tabular} & 0.469 & 0.802 & \\
VDOS + RDF + ORB descriptor & SpectraCNN & \begin{tabular}{@{}c@{}}Scalar prediction \\[-1ex] with Arrhenius-type model \end{tabular} & 0.476 & 0.806 & \\
\hline
VDOS + RDF + ORB descriptor + $D^*_{\text{5 ps}}$ & SimpleNN & \begin{tabular}{@{}c@{}}Mean-variance prediction \\[-1ex] with delta learning \end{tabular} & 0.443 & 0.816 & \\
VDOS + RDF + ORB descriptor + $D^*_{\text{5 ps}}$ & SimpleNN & \begin{tabular}{@{}c@{}}Mean-variance prediction \\[-1ex] with Arrhenius-type model \\[-1ex] with delta learning \end{tabular} & \textbf{0.398} & \textbf{0.844} & \\
\end{tabular}
\end{ruledtabular}
\end{table*}

\begin{figure*}[hbtp]
\centering
    \begin{minipage}{0.46\textwidth}
        \centering
        \includegraphics[width=\linewidth]{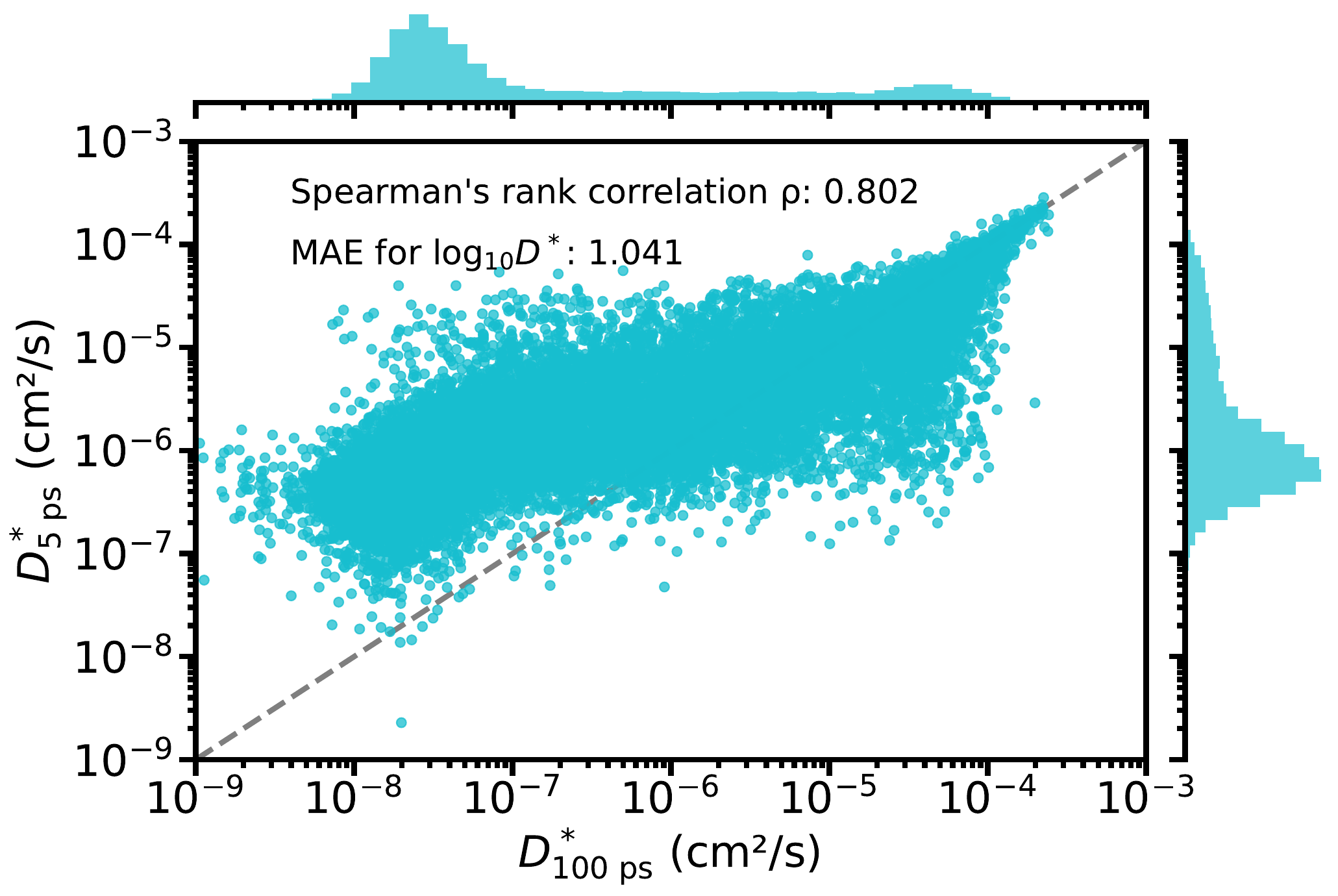}
        
    \end{minipage} \hspace{1.5em} %
    \begin{minipage}{0.46\textwidth}
        \centering
        \includegraphics[width=\linewidth]{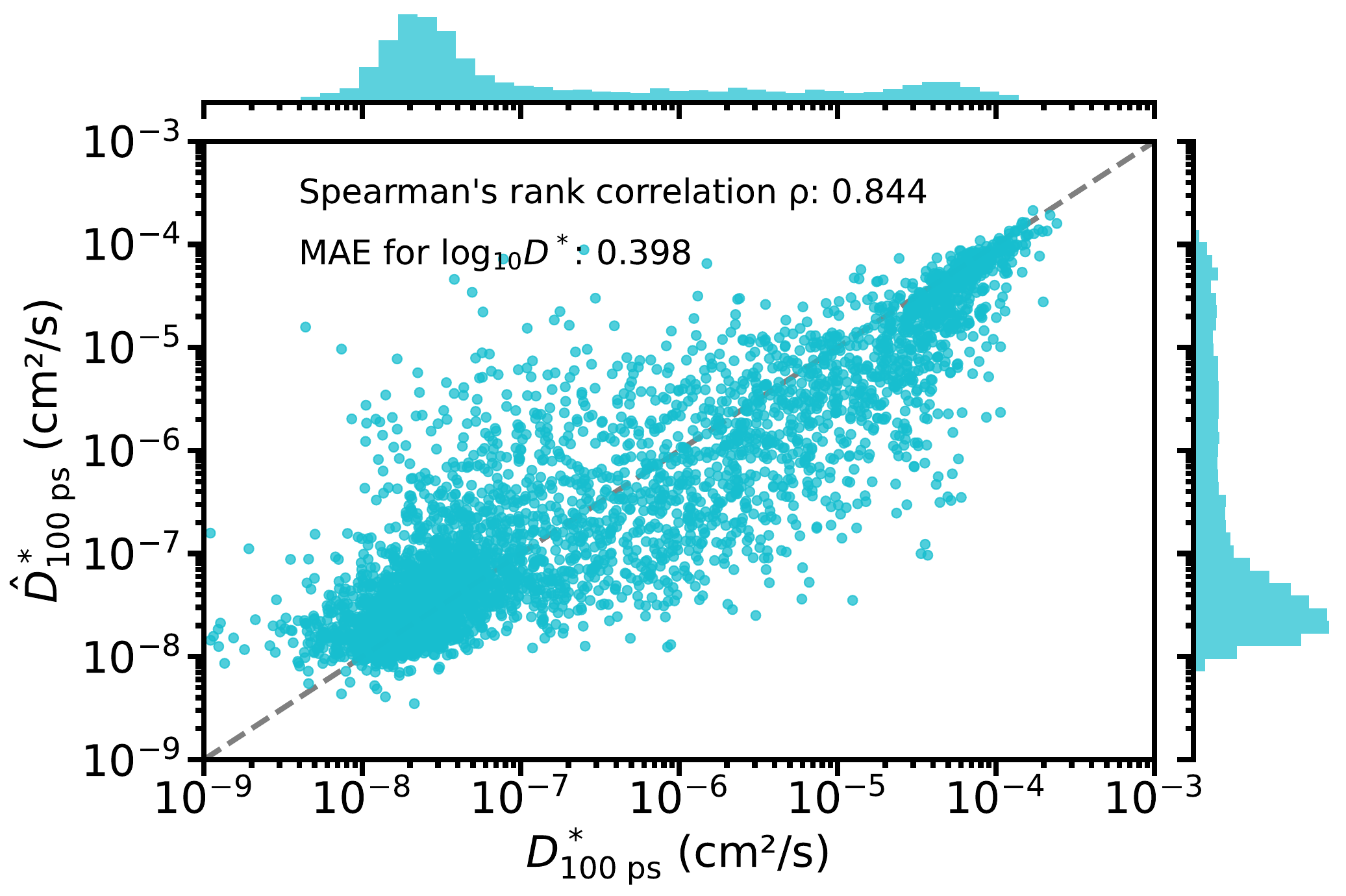}

    \end{minipage} \hspace{1.5em} %
    \\[1.5em] %
    \begin{minipage}{0.46\textwidth}
        \centering
        \includegraphics[width=\linewidth]{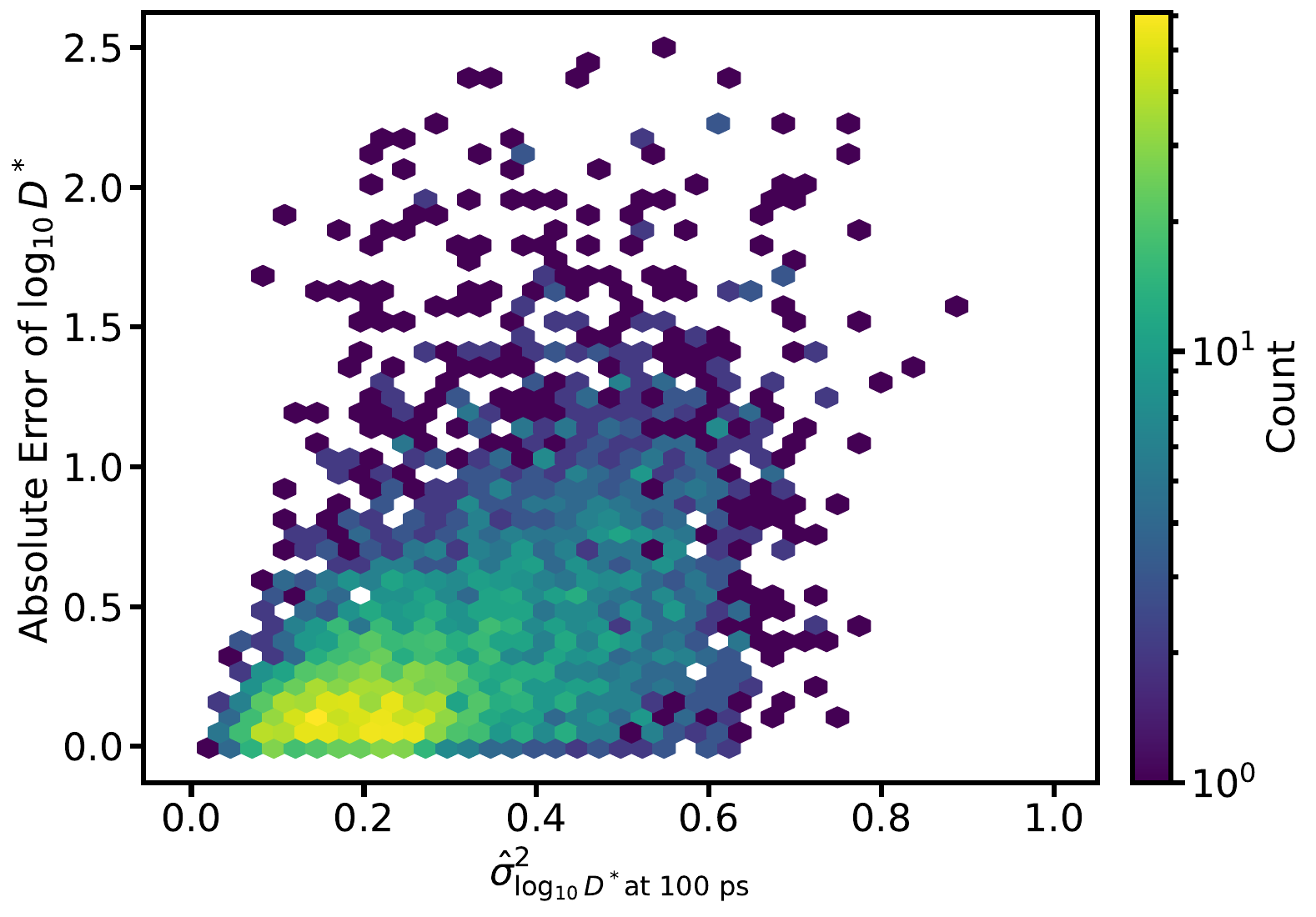}

    \end{minipage} \hspace{1.5em} %
    \begin{minipage}{0.46\textwidth}
        \centering
        \includegraphics[width=\linewidth]{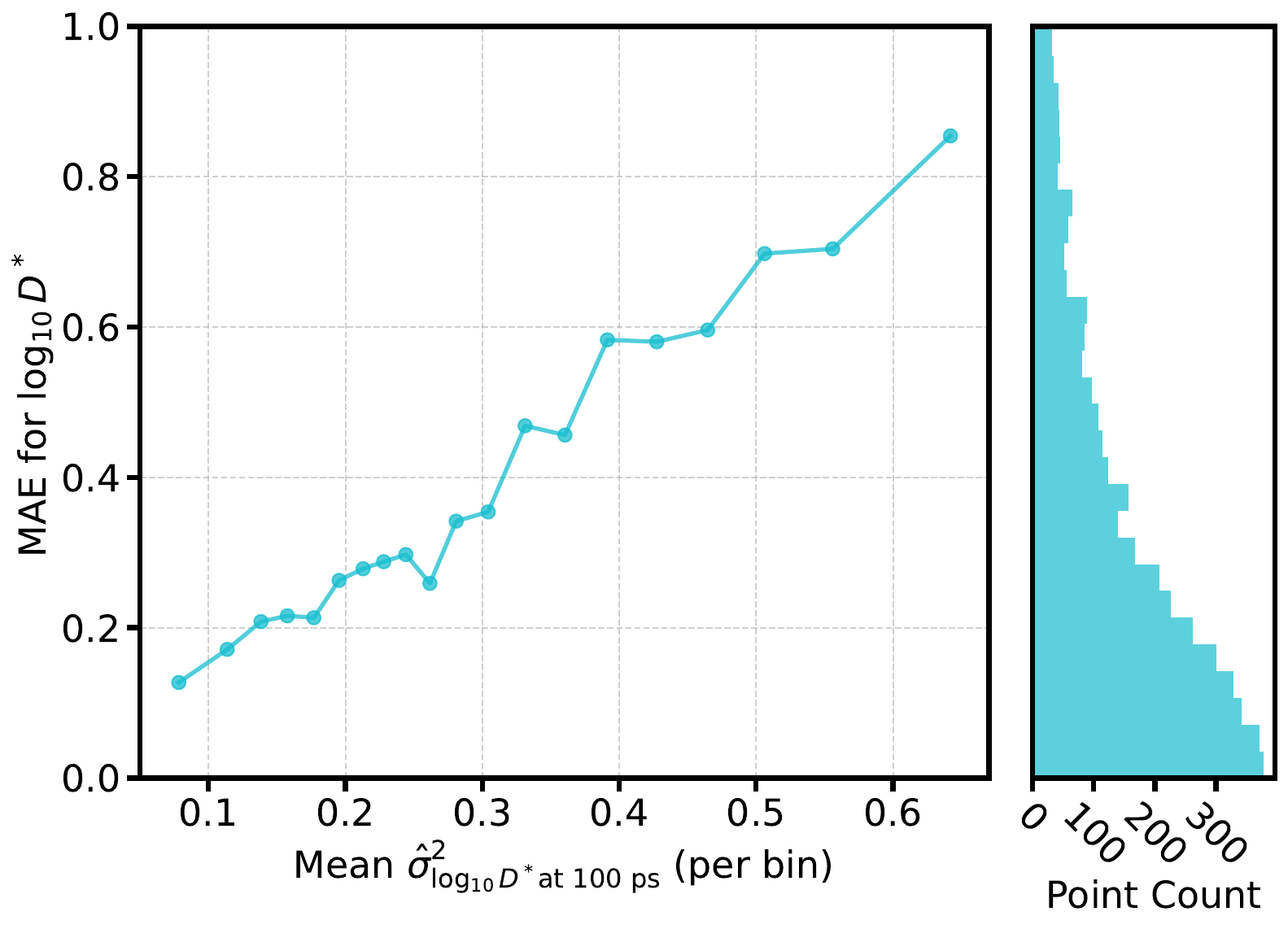}
        
    \end{minipage}

\caption{ \textit{Top left:} Parity plot of the unconverged self-diffusivity $D^*_{\text{5 ps}}$ from a 5-ps trajectory versus the converged self-diffusivity $D^*_{\text{100 ps}}$ from a 100-ps trajectory (for whole dataset: 22,290 trajectories). \textit{Top right:} Parity plot where the Arrhenius-type model with the VDOS $g(\nu)$, RDF $g(r)$, ORB embedding, and $D^*_{\text{5 ps}}$ from a 5-ps trajectory was used to predict the self-diffusivity from a 100-ps trajectory (for test set: 4,460 trajectories). The predicted self-diffusivity is represented by $\hat{D}^*_{\text{100 ps}}$ and the ground truth self-diffusivity is represented by $D^*_{\text{100 ps}}$. \textit{Bottom left:} Absolute error of $\log_{10} D^*$ versus the predicted variance $\hat{\sigma}^2_{\log_{10} D^* \text{ at 100 ps}}$ for the model shown in the top right of this figure. \textit{Bottom right:} Uncertainty calibration binning and averaging the absolute error for $\log_{10} D^*$ values versus $\hat{\sigma}^2_{\log_{10} D^* \text{ at 100 ps}}$. \label{fig:predict_diffusivity}}
\end{figure*}

As can be seen in the top left of Figure \ref{fig:predict_diffusivity}, the unconverged self-diffusivity at 5 ps ($D^*_{\text{5 ps}}$) systematically overestimates the ground-truth self-diffusivity $D^*_{\text{100 ps}}$. Predicting diffusivity at higher temperatures or for faster ion conductors is fundamentally an easier task due to the rapid convergence of diffusion statistics with many ion hopping events. However, even superionic conductors such as Li$_{10}$Ge(PS$_6$)$_2$ in the temperature range of interest for most practical applications of solid-state batteries ($-50$ to $150^{\circ}$C) have diffusivities on the order of 10$^{-9}$ to 10$^{-6}$ cm$^2$/s. While it may appear from the top left of Figure \ref{fig:predict_diffusivity} that a fast ion conductor can be identified nearly perfectly with only a 5-ps trajectory, the systematic overprediction of $D^*$ for many poor ion conductors at 5 ps results in too many false positives for $D^*_{\text{5 ps}}$ alone to be a reliable metric for distinguishing between poor ion conductors and fast ion conductors. The top right of Figure \ref{fig:predict_diffusivity} shows the parity plot for the self-diffusivity $\hat{D}^*_{\text{100 ps}}$ predicted from the fingerprint constructed solely using data from a 5-ps trajectory versus the ground truth self-diffusivity $D^*_{\text{100 ps}}$ (with the mean-variance prediction approach). This corresponds with the best-performing Arrhenius-type model from Table \ref{tab:table1}. Compared to $D^*_{\text{5 ps}}$, this model provides better predictive capability across the full range of diffusivities, retaining high prediction accuracy for fast ion conductors and preventing systematic overestimation of the diffusivity of slow ion conductors.

An additional advantage of the mean-variance prediction approach is that it provides the uncertainty estimate via the predicted heteroscedastic variance $\hat{\sigma}^2_{\log_{10} D^* \text{ at 100 ps}}$. As can be seen in the bottom row of Figure \ref{fig:predict_diffusivity}, the mean absolute error for $\log_{10} D^*$ correlates nicely with $\hat{\sigma}^2_{\log_{10} D^* \text{ at 100 ps}}$, indicating that $\hat{\sigma}^2_{\log_{10} D^* \text{ at 100 ps}}$ can serve as a well-calibrated proxy for the model's uncertainty. 

The key for identifying candidate solid electrolytes in a high-throughput manner is a fast screening method with few false negatives, particularly in the high diffusivity range. Potentially promising candidates that are pruned never make it to the next step in the funnel. Given that the next step in a conventional high-throughput screening funnel is running longer MD simulations (with MLIPs or DFT) on a subset of promising candidates prior to experimental validation, false positives are not as detrimental as false negatives. Nevertheless, false positives would still waste compute.

With this model trained on a large dataset of 100-ps MD trajectories, only short MD trajectories are now needed for performing inference and screening to get a reasonable prediction of $D^*$ for a candidate solid ion conductor.

\section{Discussion}
\label{sec:discussion}

The resulting transport properties that arise from the free energy landscape were the end goal in this work, rather than the computation of thermodynamic properties themselves. For this reason, deep learning was used to directly predict the transport properties from spectra (VDOS and RDF), skipping the steps involving: (1) explicit computation of thermodynamic properties from spectra and (2) empirical relations between thermodynamic properties and transport properties. The VDOS, which contains information related to the vibrational entropy, clearly plays the dominant role in predicting ion diffusion, while the RDF and ORB embedding provide some information regarding the configurational entropy and perhaps enthalpic migration barriers.

While prediction of $D^*$ at higher temperatures ($>$1000 K) is fundamentally an easier prediction task, practical operating temperatures for solid-state batteries are typically lower, which is why comparatively lower temperatures were targeted in this study. Since this method relies on probing the thermodynamic landscape rather than sampling ion hopping that may occur on a much longer timescale, low temperatures are where this prediction method may provide the most benefit.

While only the Li$^+$ vibrational modes and the Li$^+$-Li$^+$ structural information from MD were used for prediction in this study, the metrics presented in this work for the coupling between vibrational modes of Li$^+$ and framework ions ($\gamma^2_{\mathrm{Li^+, X^-}}(\nu)$, $\theta (\nu)$) could help attribute framework characteristics to fast ion diffusion, which may be useful for inverse design when conditioning a generative model on the vibrational spectra. With frequency-resolved force-force correlation information, the vibrational information could also capture some notion of how easily distorted the framework of the crystalline solid is around the Li$^+$ local environment.

While $s^{ex}_{2,config}$ and $s_{vib}$ appear to be sufficient for predicting ion diffusion coarsely, using more computationally-expensive representations of the 3-body and 4-body interactions contributing to $s_{config}$ may also prove useful in attribution of fast ion conduction to framework characteristics. Hong et al. recently developed a generalized methodology for computing the configurational entropy of both solid and liquid phases \cite{Hong2025ASolids}, extending beyond just the 2-body, excess configurational entropy. The angular distribution functions (ADFs) between Li$^+$-Li$^+$-Li$^+$ can capture 3-body interactions that are particularly important for correlated motion involved with multi-ion collisions that lead to migration events in the mobile ion sublattice. ADFs between Li$^+$-X$^-$-Li$^+$ or Li$^+$-X$^-$-X$^-$ may also capture lattice relaxation or librational motion.

\section{Methods}
\label{sec:methods}

\subsection{Molecular dynamics simulations of crystalline solid ion conductors}

Solid ion conductors were chosen from Materials Project \cite{Jain2013Commentary:Innovation} to run MD at 600, 700, 800, 900, and 1000 K. The solid ion conductors were selected by the following criteria: (1) more than 10\% of the atoms are lithium, (2) bandgap is $>$2 eV, and (3) energy above the convex hull is $<$0.1 eV/atom. For each of the selected 4,458 solid ion conductors, a supercell was constructed to meet a minimum dimension of 9.0 \AA{} along all of the three crystallographic axes $a$, $b$, $c$. Allowing the cell and atomic positions to change, the crystal was relaxed to a maximum force threshold of 0.01 eV/\AA{} using the ORB-v3 interatomic potential \cite{Rhodes2025Orb-v3:Scale} (\texttt{orb-v3-conservative-inf-omat} without dispersion correction, float32-high precision, as implemented in \texttt{orb 0.5.5}). MD was driven with the same ORB interatomic potential to evaluate forces and energies. MD simulations were run with the Nosé-Hoover chains thermostat in \texttt{ase 3.26.0} \cite{HjorthLarsen2017TheAtoms} (damping constant of 40 fs and timestep of 1.0 fs) in an NVT ensemble for 100 ps. Each MD simulation was initialized with a Maxwell-Boltzmann distribution at the target temperature. Diffusion statistics (both $D^*_{\text{5 ps}}$ and $D^*_{\text{100 ps}}$) were evaluated by Bayesian regression using \texttt{kinisi 2.0.0} to obtain a distribution of self-diffusivity values from MD trajectories, and the mean of this posterior distribution was used for $D^*$ in the instance of scalar prediction \cite{McCluskey2024AccurateSimulation}. The first 1 ps was discarded in computing the diffusion statistics for all simulations, so as to avoid the ballistic regime at the start of MD.

\subsection{Model and training hyperparameters}

\textit{SimpleNN}: As a baseline model, a standard feedforward neural network, or multi-layer perceptron (MLP), was implemented. The architecture consists of a single, fully-connected hidden layer with 128 neurons followed by a ReLU activation function. To handle multi-channel inputs, the 1D spectra are first flattened into a single contiguous vector before being passed through the network to predict the target scalar property. Additionally, when the network is utilized as a sub-module within a larger joint architecture, an optional final ReLU activation is applied to the output.

\textit{SpectraCNN}: It was initially hypothesized that a CNN could capture interactions between vibrational modes at different frequencies within and across channels in the 1D frequency-domain spectra shown in Figure \ref{fig:xdoss}. Given that a CNN exhibits translation invariance, it would not be suited on its own for predicting properties from spectra, so the frequency (for VDOS) or radial (for RDF) coordinate was added as an explicit spatial channel for each spectrum. A simple CNN with a single, fixed dilation and kernel size failed to accurately predict diffusivity. To capture both local peakedness and broader global characteristics, a multi-stage, multi-branch architecture was employed. The network consists of three sequential Inception-style blocks, where each block processes the input through parallel 1D convolutional branches with varying kernel sizes (3, 7, and 13) and branch-specific dilations, followed by a $1\times1$ convolution to mix the channel features. Channel-wise attention was incorporated using Squeeze-and-Excitation blocks. To systematically expand the receptive field, the dilations across the parallel branches were exponentially scaled up at each subsequent stage (e.g., base dilations of 1, 2, and 4 in the first stage, scaling up in stages two and three). To ensure local peak information was retained, features from the first stage were passed through a $1\times1$ bottleneck convolution as a long skip connection. Finally, adaptive global average pooling (GAP) and global max pooling (GMP) were performed independently on both the early skipped features and the final deep features. These pooled representations were concatenated and passed through a multi-layer perceptron readout head utilizing GELU activations and dropout (10\%) for the final scalar prediction.

\textit{Training hyperparameters}: For all models except those with the task of mean-variance prediction, training was performed for 500 epochs with a batch size of 250, using mean-squared error loss and a learning rate of 10$^{-4}$. For mean-variance prediction, the training hyperparameters were the same, but the loss function consisted of equally-weighted Kullback-Leibler divergence loss and mean-squared error loss. The target distribution was given by the mean and variance from a Gaussian fitted to the \texttt{kinisi} posterior for $\log_{10} D^*_{ \text{ 100 ps}}$, and the predicted distribution was given by the predicted mean and variance of $\log_{10} \hat{D}^*_{ \text{ 100 ps}}$. While fitting skew-normal distributions provided better fits to the target $\log_{10} D^*_{ \text{ 100 ps}}$ distributions, they empirically did not provide better performance in the diffusivity prediction task. For all models, an 80-20 train-test split was employed with the aforementioned group shuffle split by material identifiers. 


\begin{acknowledgments}
Simulations for preliminary studies and analysis were run using the computing resources of MIT SuperCloud \cite{Reuther2018InteractiveAnalysis}. This research used resources of the National Energy Research Scientific Computing Center (NERSC), a Department of Energy Office of Science User Facility (NERSC award ALCC-ERCAP-m5068). J.N. acknowledges support from the Mathworks Fellowship.

\end{acknowledgments}



\bibliography{references}

\clearpage
\onecolumngrid

\setcounter{equation}{0}
\setcounter{figure}{0}
\setcounter{table}{0}
\setcounter{page}{1}
\setcounter{section}{0}

\makeatletter
\renewcommand{\theequation}{S\arabic{equation}}
\renewcommand{\thefigure}{S\arabic{figure}}
\renewcommand{\thetable}{S\arabic{table}}
\renewcommand{\thepage}{S\arabic{page}}
\renewcommand{\thesection}{S\arabic{section}}

\renewcommand{\theHequation}{S\arabic{equation}}
\renewcommand{\theHfigure}{S\arabic{figure}}
\renewcommand{\theHtable}{S\arabic{table}}
\renewcommand{\theHsection}{S\arabic{section}}

\begin{center}
    \textbf{\large Supplementary Information for: Vibrational, structural, and chemical fingerprints of ion diffusion in crystalline solids}\\[0.5cm]
    Gavin Winter,$^1$ Juno Nam,$^1$ and Rafael G\'omez-Bombarelli$^{1,*}$\\[0.2cm]
    \textit{$^1$Department of Materials Science and Engineering, Massachusetts Institute of Technology, Cambridge, MA 02139, USA}\\[0.2cm]
    $^*$Email: rafagb@mit.edu
\end{center}
\vspace{0.5cm}

\onecolumngrid 

\begin{figure*}[hbtp]
    \centering
    \begin{minipage}{0.99\textwidth}
        \centering
        \includegraphics[width=\textwidth]{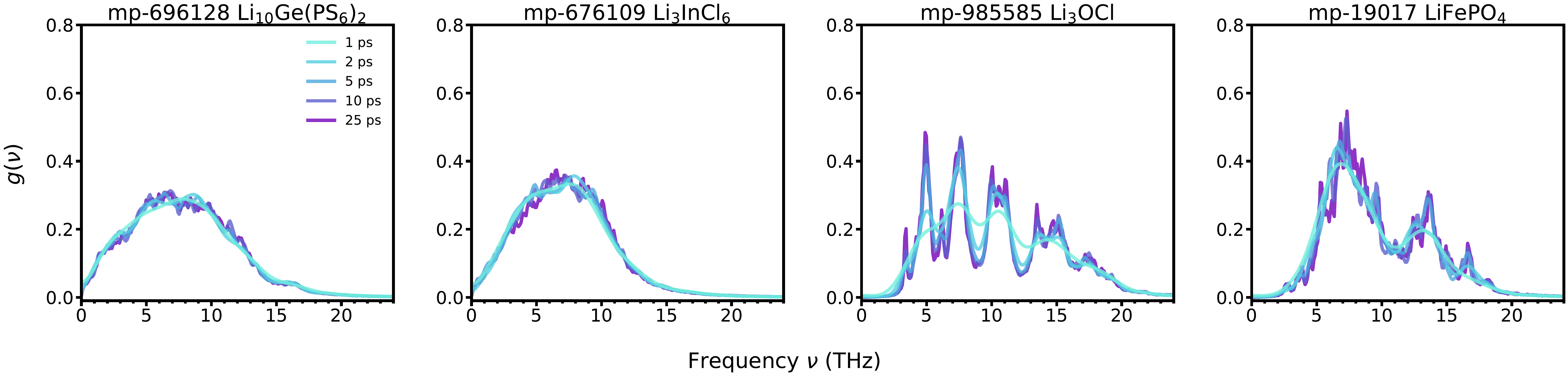}
    \end{minipage}
    \begin{minipage}{0.99\textwidth}
        \centering
        \includegraphics[width=\textwidth]{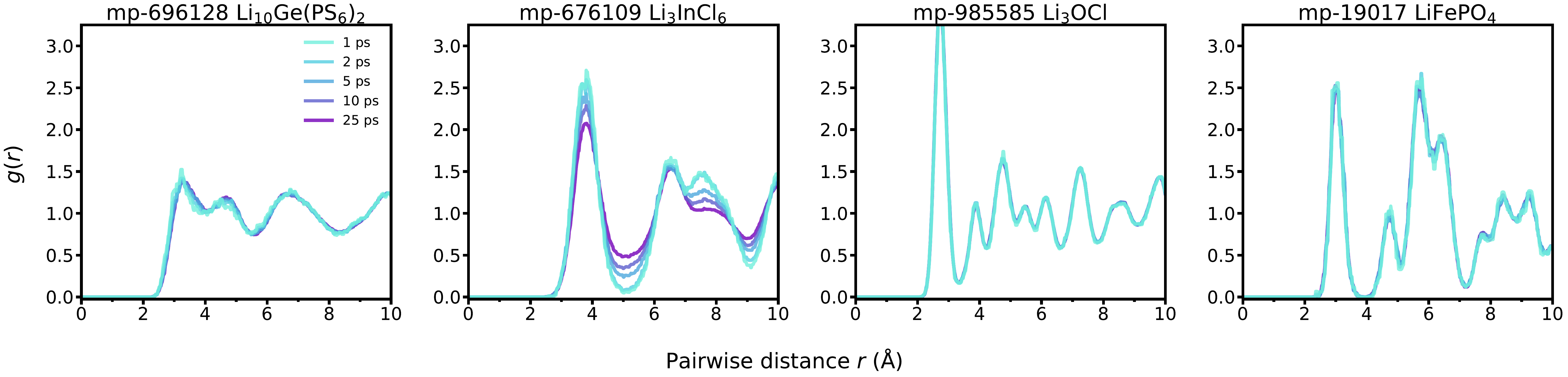}
    \end{minipage}
    \caption{ Convergence of characteristic features for representative solid ion conductors with simulation time at 600 K. \label{fig:vdos_rdf_convergence}}
\end{figure*}

Figure \ref{fig:vdos_rdf_convergence} shows the VDOS $g(\nu)$ and RDF $g(r)$ calculated at 600 K for a trajectory of 1, 2, 5, 10, and 25 ps in simulation length. While it seems like some features are not able to be resolved in the top of Figure \ref{fig:vdos_rdf_convergence}, this may be an artifact resulting from the Fourier transform being performed with segment lengths of 64, 128, 256, 512, and 1024, respectively, for the simulation times listed. These segment lengths are listed in terms of the number of frames, where a frame is saved every 10 fs. Generally, features in both the VDOS and RDF are nearly fully resolved after 5 ps.

The first raw moment $\mu_1$ and second moment $m_2$ of $g(\nu)$ are defined in the main text. The higher order moments $m_3$ and $m_4$ relate to the skewness and kurtosis of $g(\nu)$, respectively.
\begin{equation}
    \text{skewness} = \frac{m_3}{\sigma^3} = \frac{\int_0^\infty (\nu - \bar{\nu})^3 g(\nu) d\nu}{(\int_0^\infty (\nu - \bar{\nu})^2 g(\nu) d\nu)^{3/2}}
    \label{eqn:skew}
\end{equation}
\begin{equation}
    \text{kurtosis} = \frac{m_4}{\sigma^4} = \frac{\int_0^\infty (\nu - \bar{\nu})^4 g(\nu) d\nu}{(\int_0^\infty (\nu - \bar{\nu})^2 g(\nu) d\nu)^{2}}
    \label{eqn:kurtosis}
\end{equation}
The skewness quantifies the relative weighting of soft (low-$\nu$ modes) versus hard (high-$\nu$ modes). A slight right skew $g(\nu)$, corresponding with a high density of low-frequency modes and a long tail extending to high-frequency modes, should be characteristic of fast solid ion conductors that have a liquid-like Li$^+$ sublattice. The kurtosis indicates the presence of extreme outliers, such as sharp peaks at specific frequencies. Low kurtosis should be characteristic of the disordered Li$^+$ sublattice in fast ion conductors that have many different local environments each with their own distinct vibrational modes for Li$^+$ to access.

\begin{figure*}[hbtp]
    \centering
    \includegraphics[width=\textwidth]{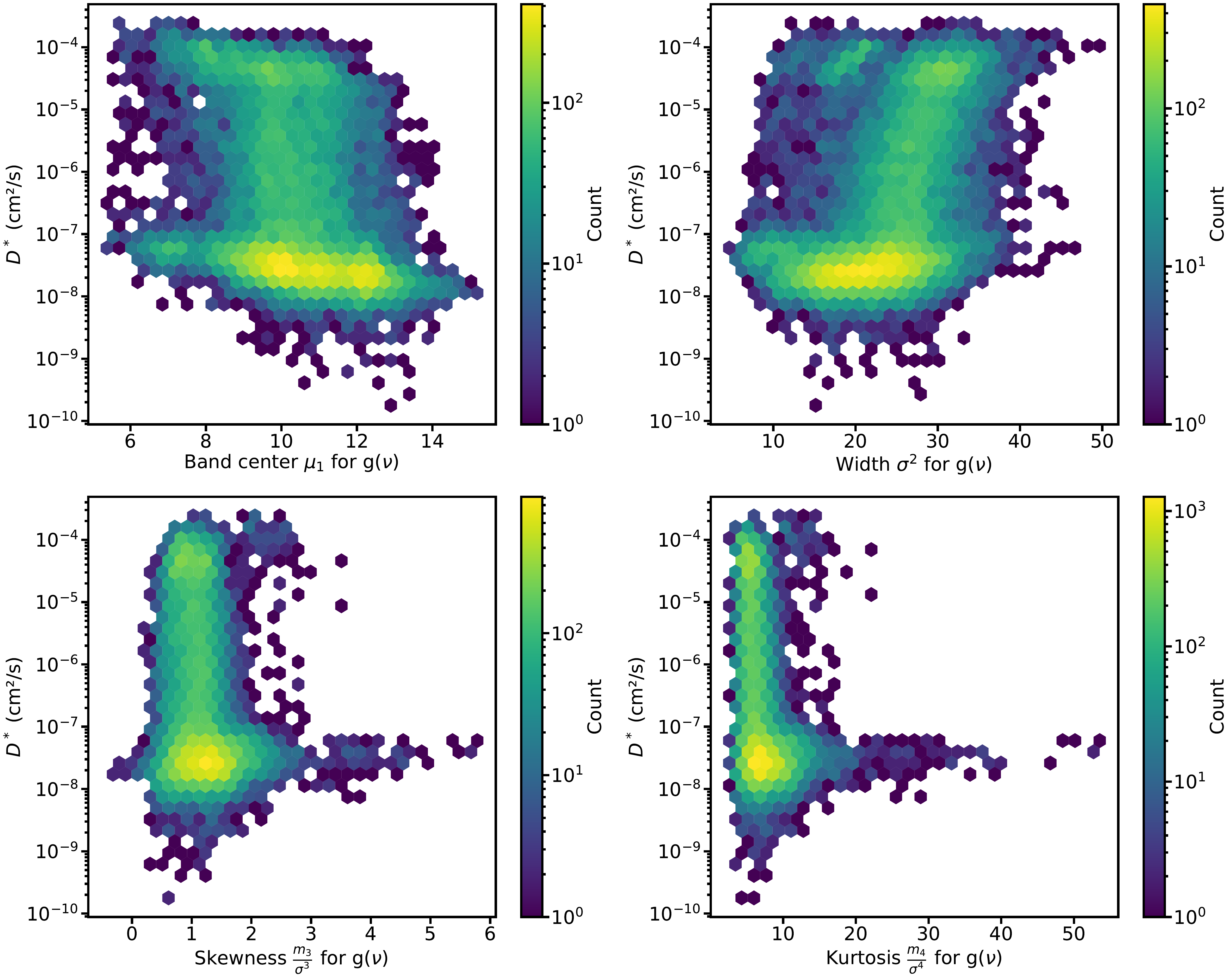}
    \caption{ Correlation of self-diffusivity with the moments of the VDOS distribution for the full set of 4,458 solid ion conductors compiled from all temperatures (600, 700, 800, 900, 1000 K). \label{fig:correlate_diffusivity_vdos_fullmoments}}
\end{figure*}

As seen in the bottom row of Figure \ref{fig:correlate_diffusivity_vdos_fullmoments}, a relatively low skewness (high symmetry in $g(\nu)$) and low kurtosis are generally needed for fast ion conduction. The skewness and kurtosis thus may provide only a weak criteria on which to exclude a candidate solid ion conductor from being a fast ion conductor, and are not sufficient metrics for correlation by themselves. 

\end{document}